\documentclass[a4paper,11pt]{article}

\usepackage{pos}
\usepackage{siunitx}
\usepackage{cleveref}
\usepackage{acro}
\usepackage{bm}
\usepackage{multirow}
\usepackage{booktabs}
\usepackage{array}
\usepackage{amsmath}
\usepackage{tablefootnote}
\usepackage{threeparttable}
\usepackage[mathlines]{lineno}
\usepackage{lipsum} 

\title{Search for Sub-Relativistic Magnetic Monopoles with the IceCube Neutrino Observatory}

\ShortTitle{Search for Sub-Relativistic Magnetic Monopoles}

\author{The IceCube Collaboration \\{\normalsize \normalfont(a complete list of authors can be found at the end of the proceedings)}\\}

\emailAdd{jonas.haeussler@rwth-aachen.de}
\emailAdd{nmoller1@icecube.wisc.edu}

\abstract{

Magnetic monopoles are beyond standard model particles, predicted by Grand Unified Theories (GUTs) to be created during the early universe. At typical masses of the GUT-scale - above $10^{14}$ GeV - these particles would move at sub-relativistic speeds. The Rubakov-Callan effect predicts that magnetic monopoles can catalyze nucleon decays, in particular the decay of protons. This results in a unique signature of small particle cascades along the trajectory of the slow moving magnetic monopole. Since 2012, a dedicated Slow-Particle Filter has been implemented in the IceCube Neutrino Observatory for the detection of magnetic monopoles. Current limits set an upper bound for the monopole flux at $\Phi_{\mathrm{90}}\leq 10^{-17}$ to $10^{-18} \mathrm{cm}^{-2}\mathrm{s}^{-1}\mathrm{sr}^{-1}$ depending on the catalysis cross section for the proton decay. A detection of the monopole flux thus requires exceptional background rejection and signal efficiency. This is accomplished using machine learning methods. In this analysis, we use a multi-level boosted decision tree classifier. We present the strategy behind the background and signal simulation, the classification efficiency, and IceCube’s projected sensitivity for the detection of sub-relativistic magnetic monopoles. 

\vspace{4mm}

{\bfseries Corresponding authors:}
Jonas Häußler$^{1*}$, 
Nicolas Moller$^{2}$\\
{$^{1}$ \itshape III. Physikalisches Institut, RWTH Aachen University}\\
{$^{2}$ \itshape Dept. of Physics and The International Center for Hadron Astrophysics, Chiba University}\\[4mm]
$^*$ Presenter
}

\ConferenceLogo{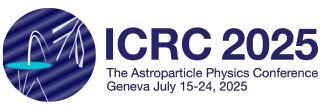}

\FullConference{39th International Cosmic Ray Conference (ICRC2025)\\
 15–24 July 2025\\
Geneva, Switzerland\\}

\DeclareAcronym{gut}{
  short = GUT,
  long  = Grand Unified Theory,
  short-plural = s,
  long-plural  = s,
}

\DeclareAcronym{guts}{
  short = GUTs,
  long  = Grand Unified Theories,
}

\DeclareAcronym{pmt}{
    short = PMT,
    long = photomultiplier tube,
    short-plural = s,
    long-plural = s
}

\DeclareAcronym{dom}{
    short = DOM,
    long = Digital Optical Module,
    short-plural = s,
    long-plural = s
}

\DeclareAcronym{slop}{
    short = SLOP,
    long = Slow-Particle
}

\DeclareAcronym{bdt}{
    short = BDT,
    long = Boosted Decision Tree,
    short-plural = s,
    long-plural = s
}

\DeclareAcronym{hlc}{
    short = HLC,
    long = high local coincidence, 
    short-plural = s,
    long-plural = s
}

\DeclareAcronym{slc}{
    short = SLC,
    long = soft local coincidence, 
    short-plural = s,
    long-plural = s
}

\DeclareAcronym{frt}{
    short = FRT,
    long = Fixed-Rate-Trigger
}

\begin{document}

\maketitle

\section{Introduction}\label{sec:Introduction}

\begin{figure}[b!]
    \centering
    \includegraphics[height=0.4\linewidth]{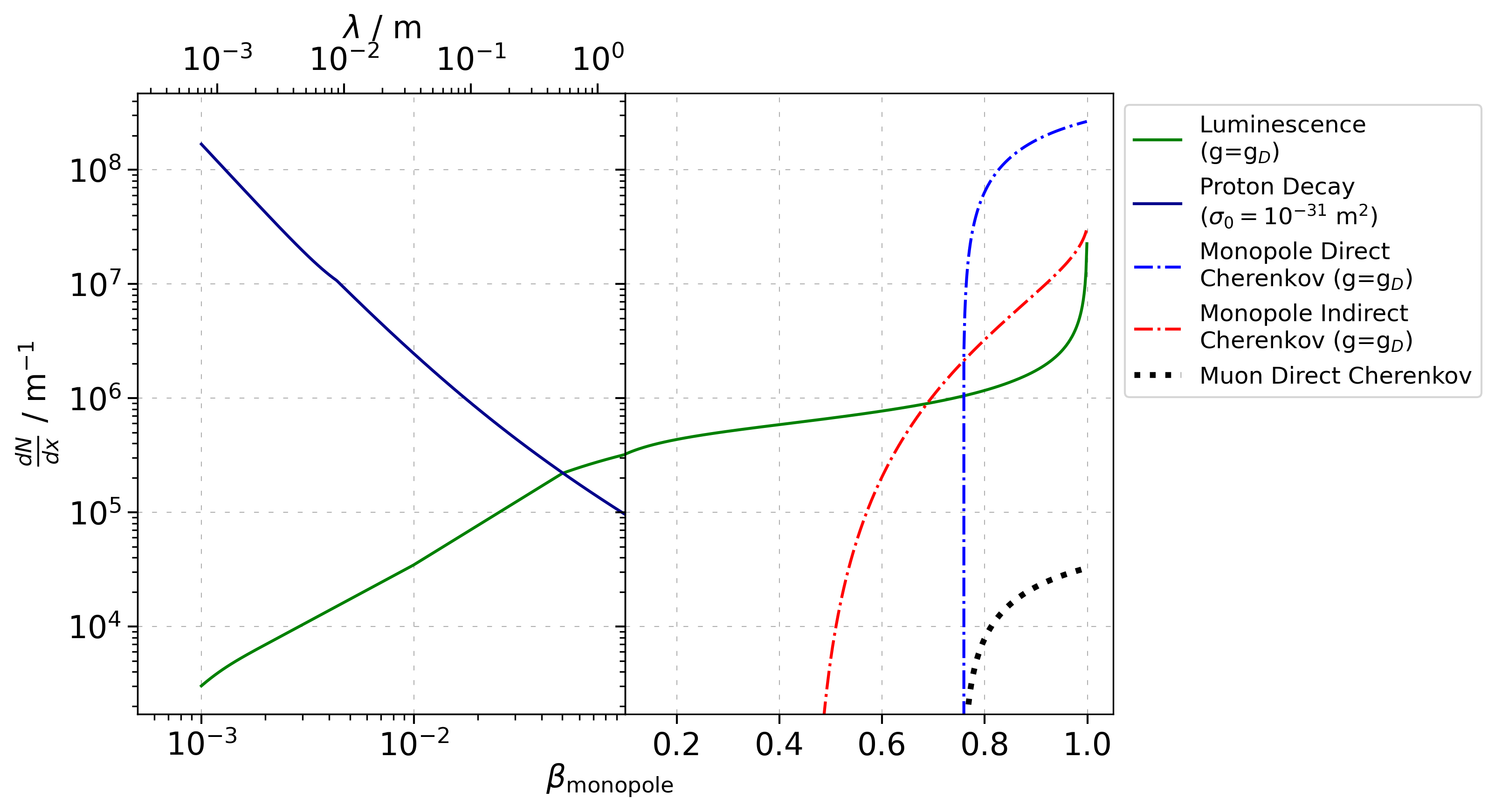}
    \caption{Light yield of magnetic monopoles with Dirac charge $g_D$ in dependence of their velocity in South Pole ice ($\rho =\SI{0.917}{\g/ \cm^{3}}$). Shown are the contributions due to Luminescence, Cherenkov light, and from proton decays induced by the Rubakov-Callan effect, with a mean free path $\lambda$ depending on the velocity and underlying cross-section. For comparison, the light yield of a muon via direct Cherenkov radiation is shown as a black dotted line.}\label{fig:light_yield}
\end{figure}

Maxwell's equations for the electromagnetic fields have two inconsistencies: they are not symmetric for electric and magnetic fields, and the quantisation of the electric charge in the form of the electron is arbitrary.
Nearly a century ago, Dirac \cite{Dirac:1931kp} showed that the introduction of a magnetic monopole --- a particle with a discrete magnetic charge that functions as a source and sink for magnetic fields --- not only symmetrises Maxwell's equations but also necessitates the quantisation of the electron.
He showed that the charge has to be a multiple of the Dirac charge
\begin{equation}
  g_{\mathrm{D}} = \frac{e}{2\alpha}\,,
\end{equation} 
where $e$ is the elementary electric charge and $\alpha$ is the fine structure constant.
't Hooft \cite{tHooft:1974kcl} and Polyakov \cite{Polyakov:1974ek} extended this by showing that during the spontaneous breakdown of a gauge group to a subgroup, a classical solution exists that leads to a soliton functioning as a magnetic monopole.
This includes the transition from a \ac{gut} gauge group to the electromagnetic $U(1)$ gauge group.
The mass of such as \ac{gut} monopole depends heavily on the specific underlying \ac{gut} gauge group and can range from between $10^{5}$ \si{\giga\eV} to $10^{18}$ \si{\giga\eV} \cite{IceCube:2014xnp}.
Masses of up to $10^{13}$ \si{\giga\eV} are called intermediate mass monopoles and are created in intermediate symmetry breaking steps below the \ac{gut} scale.
Heavy monopoles with masses on the \ac{gut} scale could be created during phase transitions in the early universe shortly after the Big Bang \cite{ParticleDataGroup:2024cfk}.
These relic monopoles could persist as stable particles up to today and are the focus of this analysis.

During their transit through the universe, these monopoles would be accelerated by galactic and extragalactic magnetic fields.
Since the acceleration of magnetic monopoles consumes energy, an upper bound on the monopole flux can be determined by requiring that the rate of this energy drain is small compared to time scales on which cosmic galactic fields can be regenerated.
This Parker \cite{Turner:1982ag} bound constrains the flux below $10^{-15}$ \si{\per\centi\meter\squared\per\second\per\steradian}.
From the observation of cosmic magnetic fields, it can be determined that the maximum transferable kinetic energy is $\sim10^{14}$ \si{\giga\eV} \cite{Wick:2000yc}, meaning that intermediate mass monopole ($m_M \leq 10^{13}$ \si{\giga\eV}) travel at relativistic speeds, while heavy monopoles travel at sub-relativistic speeds.

This has consequences for the light yield of magnetic monopoles, as depending on their velocity, they interact differently with their surrounding medium.
\Cref{fig:light_yield} shows for monopoles of different velocities $\beta$ the light yield of different processes in ice.
For relativistic monopoles, Cherenkov light is the dominant contribution to the emitted light.
This can be either direct Cherenkov light emitted by the monopole during transit or indirect Cherenkov light emitted by $\delta$-electrons, which are kicked off their bond with an atom by the monopole.
Additionally, ionizing particles can excite atomic electrons, producing luminescence light \cite{Pollmann:2021jlo} when the later de-excite after a short delay. This process dominates at intermediate velocities ($0.1\lesssim \beta \lesssim 0.6$).

However, for this analysis, the most relevant light production mechanism is the nucleon decay via the Rubakov-Callan effect \cite{Rubakov:1982fp,Callan:1982au}.
At masses above $10^{14}$ \si{\giga\eV}, the \ac{gut} gauge bosons at the centre of the soliton allow for baryon number violation, which means that monopoles can catalyse the decay of nucleons, for example, the proton decay $M + p \rightarrow M + e^{+} + \pi^{0}$.
Theoretically, other nucleon decays are also viable; however, in the proton decay, nearly the entire rest mass of the proton is converted into electromagnetic particles, which form electromagnetic cascades.
As such, this process has a very high light yield and is used as a benchmark in this analysis.

The light produced by monopoles can be measured by conventional particle detectors.
This analysis uses the cubic-kilometre-sized IceCube Neutrino Observatory \cite{Aartsen:2016nxy}, located at the South Pole.
It consists of 5160 \acp{dom} arranged along 86 strings at depths ranging from \SI{1.5}{\km} to \SI{2.5}{\km} below the Antarctic ice.
Each \ac{dom} consists of a large hemispherical \ac{pmt} which measures the Cherenkov light emitted by relativistic particles, for example muons from neutrino interactions.
Sub-relativistic magnetic monopoles have a unique signature in this detector: along their track, the slow particles produce small electromagnetic cascades with a mean free path $\lambda$ which depends on the Rubakov-Callan cross-section $\sigma_0$.
No other standard model particle is expected to have such a signature.

For the search of sub-relativistic particles in IceCube, including magnetic monopoles, a dedicated trigger was conceived and deployed in 2012 (more details in the following section).
The \ac{slop} trigger was used in a previous analysis by IceCube \cite{IceCube:2014xnp} to set the current world-leading limits, which constrain the flux of sub-relativistic magnetic monopoles down to a level of $\Phi_{90} \leq 10^{-18} (10^{-17})$ \si{\per\centi\meter\squared\per\second\per\steradian} for catalysis cross sections of $10^{-22} (10^{-24})$ \si{\centi\meter\squared} at a confidence level of 90\%.
This limit was achieved with a livetime of one year and with the \ac{slop}-trigger running only on DeepCore, a denser infill area of IceCube.
This analysis extends the previous analysis by applying modern machine learning methods for the classification of monopoles and utilizing the full 12 years of \ac{slop}-data running on the entire detector.

\section{SLOP-Trigger}\label{sec:SLOP}

\begin{figure}
    \centering
    \begin{minipage}[c]{0.4\textwidth}
      \includegraphics[width=\textwidth]{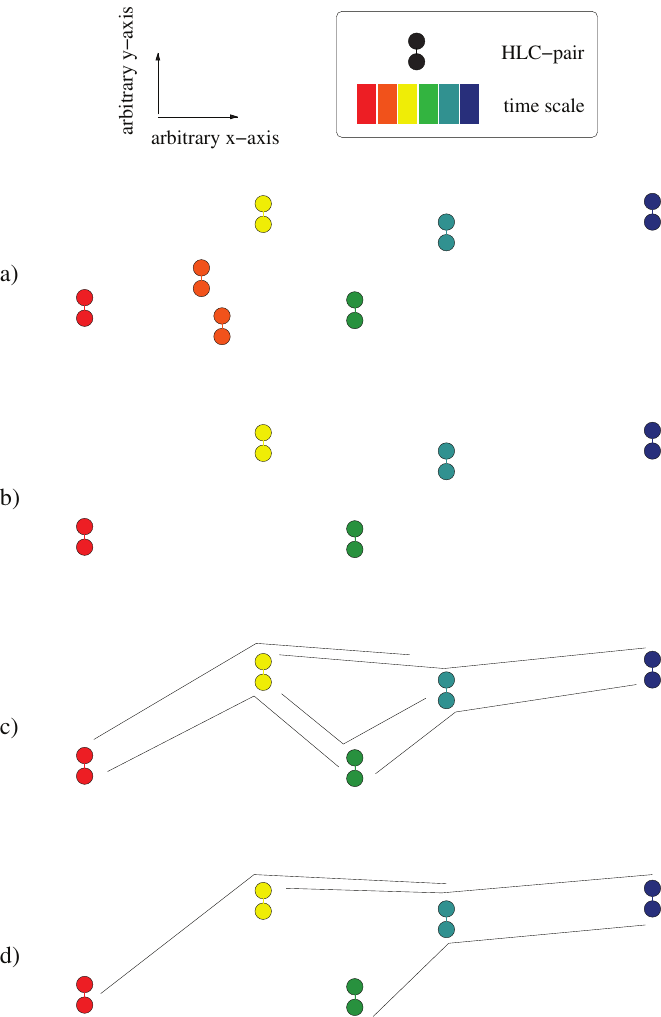}
    \end{minipage}\hspace{1cm}
    \begin{minipage}[c]{0.4\textwidth}
      \caption{
         Visualisation of the \acs*{slop}-Trigger. In the first step, the \acs*{hlc} pairs of an event are selected. Each \acs*{hlc} pair consists of two neighbouring hits. Any pairs, which follow fast after one another (in this example the orange \acs*{hlc}-pairs) are most likely induced by a relativistic particle, and are removed in the second step. From the remaining pairs, all possible triplets, which do not exceed a maximum timespan between two consecutive pairs, are formed. In a last step, it is checked which of these triplets are consistent with a slow particle, i.e. checking that the triplet is along a straight line, and is compatible with a constant velocity. A trigger is formed if of these valid triplets, at least 5 overlap in time. Taken from \cite{IceCube:2014xnp}.
      } \label{fig:SLOP}
    \end{minipage}
\end{figure}

The \acs*{slop}-Trigger was implemented for the full detector configuration in 2012.
For this analysis, we use 12 years of livetime up till 2024.
The \acs*{slop}-Trigger functions by looking for hits that are consistent with the profile expected from a slow particle.
The trigger continuously reads in all \ac{hlc} hits; an \ac{hlc} hit consists of two neighbouring \acp{dom} measuring a signal within a very short timespan.
By requiring two \ac{hlc} hits to have a time difference greater than a set minimum, hits induced by relativistic particles, such as muons, are suppressed.
From these selected hits, triplets are constructed.
A triplet consists of three hits, where each consecutive hit has a maximum time difference of \SI{0.5}{\ms} between one another.
These triplets are required to be along a straight line, and for the relative time difference between the hits to be consistent with a monopole of near-constant velocity.
This is to ensure that the triplets are consistent with the track of a slow particle and are not induced by random noise.
For the trigger to successfully fire, it is then required that at least 5 of these triplets are overlapping in time and are measured within a maximum event duration of \SI{5}{\ms}.
A step-by-step visualization can be seen in \cref{fig:SLOP}, and more details, including specific parameters, can be found in \cite{Latseva:2023master}.

\begin{figure}[t!]
\centering
\includegraphics[height=0.4\linewidth]{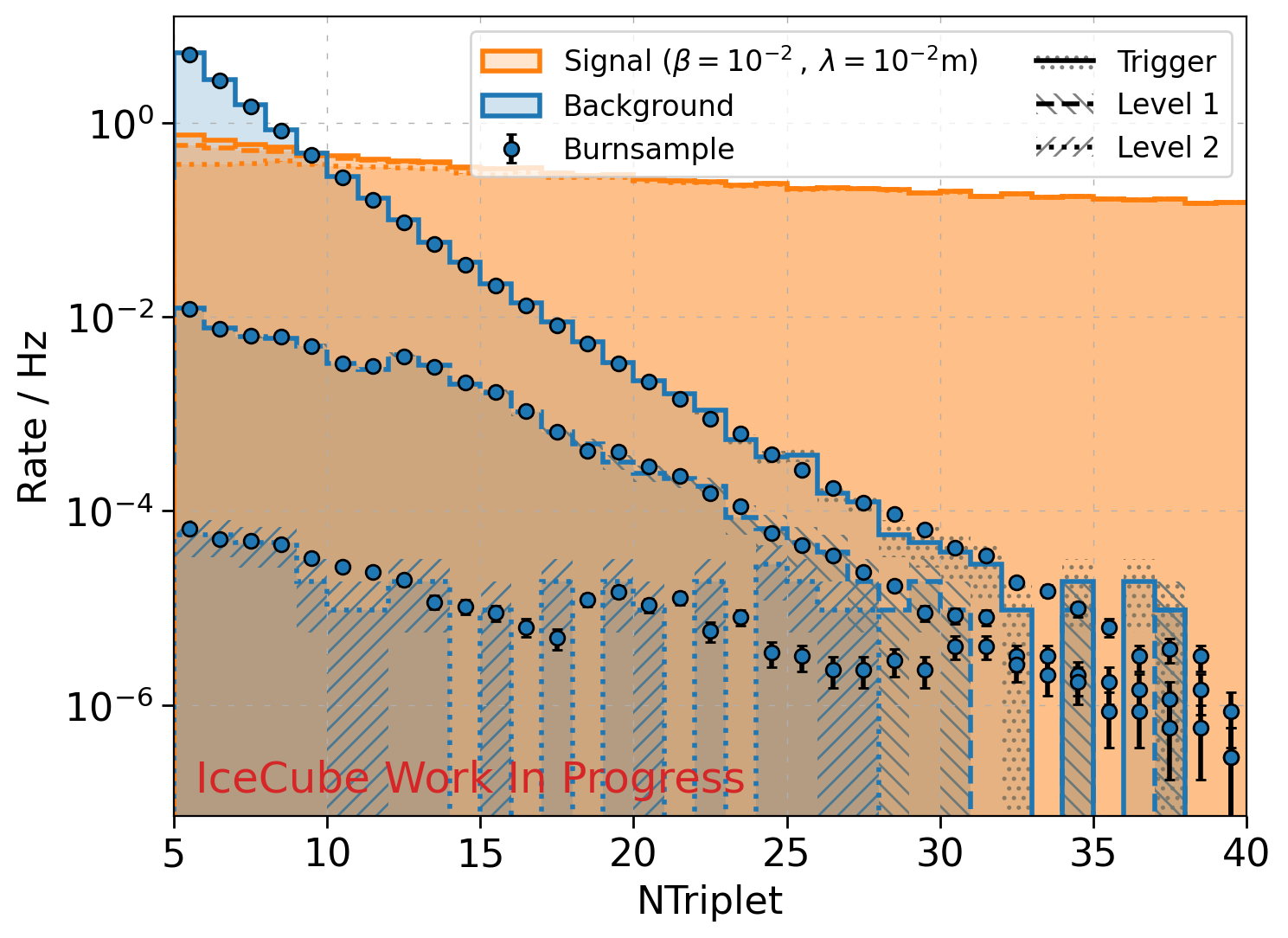}
\caption{Number of triplets per event for simulated background and signal, as well as for 1\% of detector data (burnsample). The distributions after the SLOP-Trigger are shown as solid lines, while the dashed and dotted lines are the distributions after applying the Level-1 and Level-2 BDT cuts, respectively. The \acsp*{bdt} used for the cut was trained on the same monopole parameters as the shown signal. The statistical uncertainty for the simulated data is given by the hatched region. The signal distribution extends to larger values (NTriplet $\geq 250$), which depends on the respective monopole type.}\label{fig:NTriplet}
\end{figure}

\Cref{fig:NTriplet} shows the distribution of the number of triplets in a triggered event for signal, background, and burnsample (1\% of measured detector data used for validation).
It can be seen that the background is constrained to smaller values, while the signal extends to large values depending on the specific parameters of the simulated magnetic monopole.
Three main sources of background can pass the \ac{slop} trigger: low energetic muons induced by cosmic rays resulting in single \ac{hlc} hits, late after-pulses from ionized gas in the \ac{pmt}, and dark noise (hits in absence of signal) consisting of uncorrelated noise due to thermal emission and correlated noise likely due to radioactive decays in the glass-sphere of each \ac{dom}.
The combined background results in a trigger rate of $\sim \SI{11}{\Hz}$.
More details can be found in \cite{Böttcher:2019master}.

\section{Event Selection}\label{sec:EventSelection}

For the event selection, we use a multi-level \acl{bdt} approach.
A \ac{bdt} is a popular and efficient method for classification and regression tasks, which is commonly used in the identification of particles.
For the classification, we use $36$ variables, determined from triplets passing the \ac{slop} trigger.
An overview of these variables can be seen in \cref{tab:BDT_Variables}, consisting of three basic categories: summary variables describing the event, variables describing the triplets contained in the event, and variables from a simple reconstruction.
For a detailed and complete description of these separating variables, see \cite{Latseva:2023master}.

\begin{table}[b!]
\centering
  \caption{Overview of the separating variables given to the \acp{bdt}. Statistical derivations from these quantities, such as mean and variance, are not listed. In total 36 separating variables are used. A full list and detailed explanation can be found in \cite{Latseva:2023master}.}
  \begin{threeparttable}
  \refstepcounter{table}
  \label{tab:BDT_Variables}

  \small
  \renewcommand{\arraystretch}{0.95}
  \begin{tabular}{l|l} 
  \toprule
  \textbf{Variables}            & \textbf{Explanation}                                                          \\ 
  \hline
  Trigger Length                & Duration of the event                                                         \\
  Number of triplets            & Number of triplets contained in the event                                     \\
  Number of tuples              & Number of \ac{hlc}-pairs that are part of a triplet                           \\
  Triplet to \ac{hlc} ratio     & \#Triplets/\#\acp{hlc}                                                        \\
  Smoothness\tnote{1}           & Smoothness parameter                                                          \\
  Angle\tnote{1}                & Deviation of the individual triplet directions from the mean event direction  \\
  Gap\tnote{1}                  & Spatial gap between two consecutive \ac{hlc} pairs in the hit list            \\
  Inverse velocity\tnote{1}     & Relative inverse velocity of the triplets                                     \\
  $\cos$(inner angle)\tnote{1}  & Inner angle of the triplets                                                   \\
  Velocity                      & Monopole velocity as predicted by the LineFit\tnote{2}                        \\
  $\chi^2$                      & $\chi^2$ of the LineFit\tnote{2}                                              \\
  \bottomrule
  \end{tabular}
  \begin{tablenotes}
    \footnotesize
    \item [1] Includes also stat. derived quantities such as mean, variance, quantiles, etc.
    \item [2] LineFit is a standard IceCube reconstruction that performs a simple line fit.
  \end{tablenotes}
\end{threeparttable}
\end{table}

In total, we train three levels, where each \ac{bdt} is trained on the events passing the previous level.
For the first two levels, we require the \ac{bdt} score to be $\geq0.5$ to pass the specific level.
The cut-value for the final level is chosen to maximise the sensitivity of the analysis.
Currently, the first two levels have been trained, while the training and development of the final level is ongoing.
The reason for this multi-level approach is that by only training on the output of the previous \ac{bdt} we reduce the storage requirements for the training, which allows us to focus the \ac{bdt} on rare events, training it to pick up on very small differences.

For the training and evaluation, we require extensive datasets of simulated background and signal.
Due to the long event length of a \ac{slop} event ($\mathcal{O}(\si{\ms})$) in comparison to, for example, an air shower or neutrino event ($\mathcal{O}(\si{\micro\s})$), standard procedures for simulating background data are too computionally expensive.
Instead, a data-driven approach is used.
In IceCube, every \SI{5}{\min} and for a duration of \SI{10}{\ms}, the entire detector is read out, i.e. all hits are saved.
This \ac{frt} gives a good representation of the detector noise and is used in our data-driven background simulation.

Each \ac{frt} event is cut into \textit{snips} with a length of \SI{40}{\micro\s}, which are then reshuffled.
The snip length is chosen to be much larger than any typical muon event duration, but much shorter than any monopole event, such that they are cut apart \cite{Böttcher:2019master}.
To prevent any issues from a set snip length, the snip length is smeared with an exponentially modified Gaussian distribution, with a smearing factor of 1\% of the snip length.
When recombining the snips, to avoid edge effects from not simulating a continuous data stream, we add a buffer region of \SI{5}{\ms} (the maximum duration of a \ac{slop} event) on both ends of a single simulated frame.
Additionally, we recombine \num{50000} snips ($\approx \SI{2}{\s}$) and trigger multiple events to most closely approximate a continuous data stream.

\begin{figure}[t]
\centering
\includegraphics[width=0.45\linewidth]{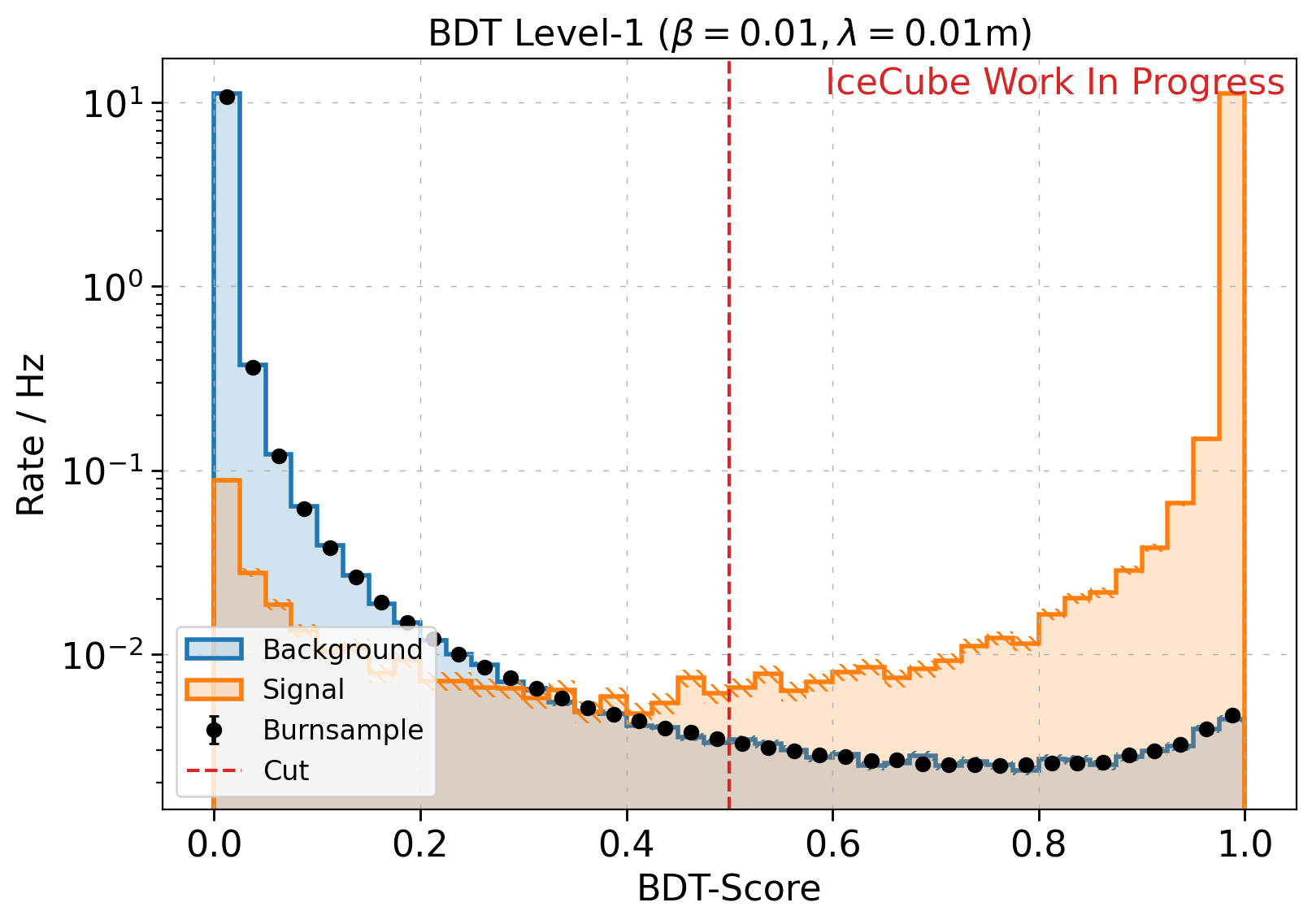}
\includegraphics[width=0.45\linewidth]{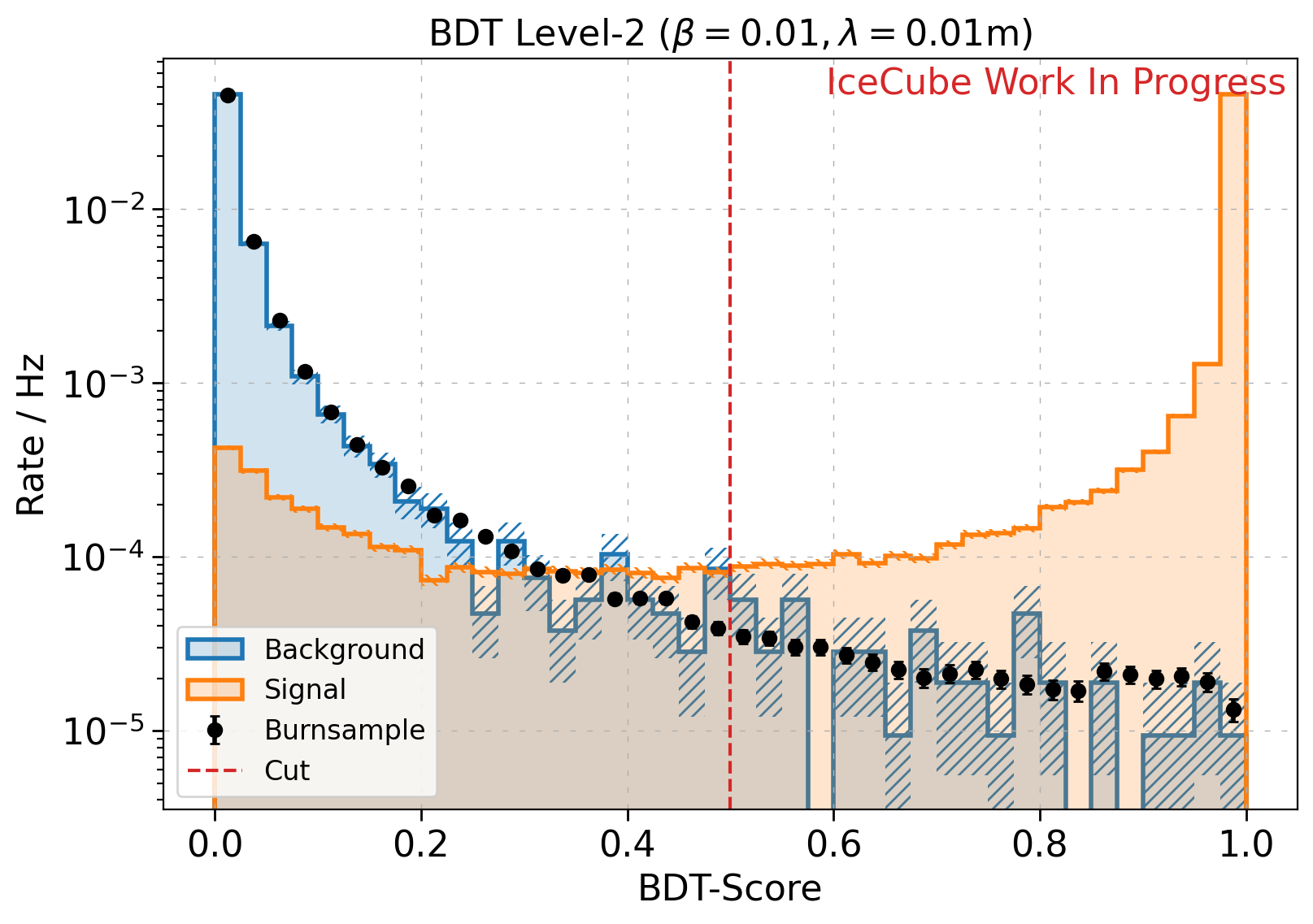}
\caption{BDT score distribution for monopole type $\beta= 10^{-2}\,,\;\lambda = 10^{-2}$ \si{\m} for Level-1 (left) and Level-2 (right). The red dashed line marks the chosen cut on the BDT-score at 0.5. In the right plot, only events which also passed the BDT Level-1 cut are shown. The signal has been normalised to the background rate.}\label{fig:BDT_Score}
\end{figure}

During the reshuffling of the snips, the correlated noise gets decorrelated.
To account for this, additional hits are added after reshuffling.
By comparing reshuffeled data with \ac{frt} data, the probability to add additional hits can be determined.
For a detailed explanation on the background simulation, see \cite{Böttcher:2019master}.

For the signal, a monopole with specific velocity $\beta$ and mean free path $\lambda$ is generated at a random point on a disk with random orientation to simulate an isotropic flux.
The monopole is then propagated through the ice, and cascades with an energy of $m_p =\SI{938}{\mega\eV}$ are simulated along its trajectory in accordance to the mean free path.
The photons from these cascades are propagated through the ice, and a full detector simulation is performed, including afterpulses in the \acp{pmt} and electronics simulation.
Additionally, untriggered background noise, as described above, is added.
More details on this can be found in \cite{Wolf:2023master}.

Currently, eight combinations of monopoles are simulated, with velocities ($\beta$) of $10^{-2}$ and $10^{-3}$ and mean free path ($\lambda$) of $10^{-2}$ \si{\m}, $10^{-1}$ \si{\m}, $1$ \si{\m} and $10$ \si{m}.
For each combination of $\lambda$ and $\beta$, a separate \ac{bdt} chain is trained.
In \cref{fig:NTriplet}, the number of triplets are shown after trigger level, and after applying the first two BDT levels trained on the monopole type $\beta=10^{-2}$ and $\lambda=10^{-2}$ \si{m}.
It can be seen that the \acp{bdt} are incredibly efficient at reducing the background while selecting nearly all of the signal.
This can also be seen in \cref{fig:BDT_Score}, showing the score for the first two levels.
For most of the trained combinations, the background rejection after Level-2 is $\geq 99.9\%$ with a combined signal efficiency $\varepsilon_{\mathrm{total}}\geq80\%$.

For the final level, we are currently in the process of simulating sufficent training datasets and developing additional variables, with a focus on geometric variables describing the position of the event in the detector, as this information is currently missing.
From preliminary studies \cite{Latseva:2023master}, we know that this level is necessary to fully reject the remaining background that can be seen in the right plot of \cref{fig:BDT_Score}.
The signal efficiency between Level-2 and Level-3 is on average $\varepsilon\sim90\%$ \cite{Latseva:2023master}, however, this is a conservative estimate, as we anticipate that with the additional seperating variables, we can further increase the perfomance of the final \ac{bdt}.

\section{Preliminary Sensitivity \& Discussion}\label{sec:Sensitivity}

\begin{figure}[t!]
\centering
\includegraphics[height=0.4\linewidth]{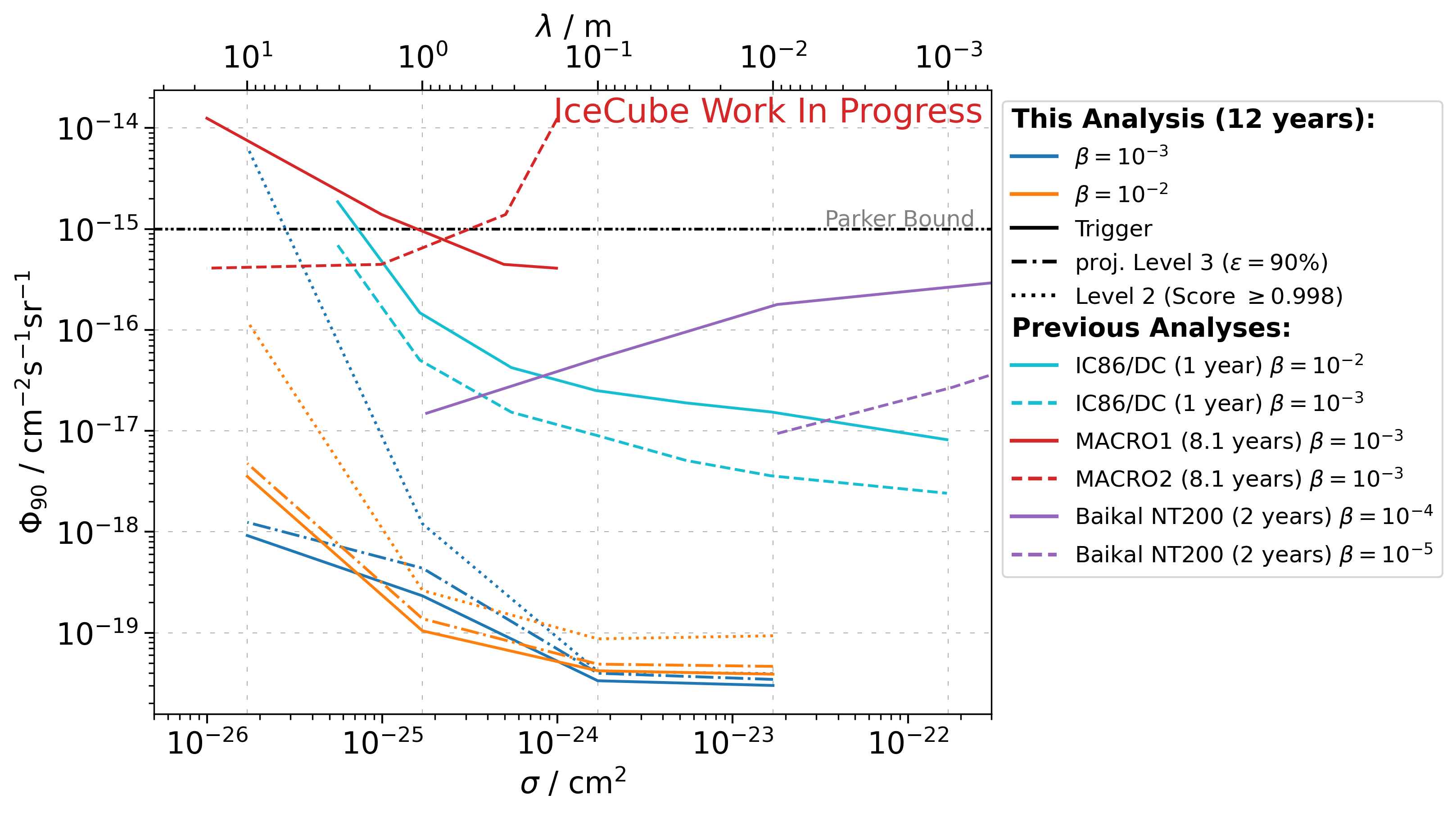}
\caption{Projected sensitivity of the final BDT selection. Shown are, under the assumption of no measured background, the sensitivities at trigger level, for Level-2 with a cut on the BDT score at 0.998, and the projected final Level-3 of the BDT selection. For the projected Level-3 sensitivity, a signal efficiency of $\varepsilon=90\%$ with respect to the standard Level-2 selection (cut at 0.5) is used, motivated by preliminary studies \cite{Latseva:2023master}. The sensitivities at trigger and Level-2 give an upper and lower bound for the Level-3 sensitivity. For comparison also the upper limits from the previous IceCube Analysis \cite{IceCube:2014xnp}, MACRO \cite{MACRO:2002iaq} and Baikal \cite{Gaponenko:2021jag} are shown.}\label{fig:Sensitivity}
\end{figure}

To quantify the improvement we can expect from this analysis in comparison to previous analyses, we determined a projected Level-3 with an assumed $\varepsilon = 90\%$.
The sensitivity for this projected Level-3 is shown in \cref{fig:Sensitivity} under the assumption that the background is completely rejected.
As upper and lower bounds on this projected sensitivity, we also included the best possible sensitivity if at trigger level we were able to completely seperate background from signal, and as a lower bound the sensitivity if we do not train the final level and instead cut on the Level-2 \ac{bdt} score at \num{0.998} to remove the background.
Comparing the projected sensitivity at $\varepsilon=90\%$ with the previous IceCube analysis shows the significant improvement of this analysis.
The main improvements are the increased livetime from \num{1} year to \num{12} years, but also the improved selection using \acp{bdt}, and the triggering on the entire detector instead of only on the DeepCore subarray.
Comparing the sensitivity at a catalysis cross section of $10^{-24}$ \si{\cm\squared} for $\beta = 10^{-2} \left(10^{-3}\right)$ shows an increase in sensitivity by a factor of $10^{3} (10^{2})$.
This analysis is expected to surpass previous limits by IceCube and other experiments in the examined parameter space and to be sensitive to monopole fluxes orders of magnitudes below current limits.


\providecommand{\href}[2]{#2}\begingroup\raggedright\endgroup

\clearpage

\section*{Full Author List: IceCube Collaboration}

\scriptsize
\noindent
R. Abbasi$^{16}$,
M. Ackermann$^{63}$,
J. Adams$^{17}$,
S. K. Agarwalla$^{39,\: {\rm a}}$,
J. A. Aguilar$^{10}$,
M. Ahlers$^{21}$,
J.M. Alameddine$^{22}$,
S. Ali$^{35}$,
N. M. Amin$^{43}$,
K. Andeen$^{41}$
% ...
\\
\\
% ...
$^{10}$ Universit{\'e} Libre de Bruxelles, Science Faculty CP230, B-1050 Brussels, Belgium \\
% ...
$^{16}$ Department of Physics, Loyola University Chicago, Chicago, IL 60660, USA \\
$^{17}$ Dept. of Physics and Astronomy, University of Canterbury, Private Bag 4800, Christchurch, New Zealand \\
% ...
$^{21}$ Niels Bohr Institute, University of Copenhagen, DK-2100 Copenhagen, Denmark \\
$^{22}$ Dept. of Physics, TU Dortmund University, D-44221 Dortmund, Germany \\
% ...
$^{35}$ Dept. of Physics and Astronomy, University of Kansas, Lawrence, KS 66045, USA \\
% ...
$^{39}$ Dept. of Physics and Wisconsin IceCube Particle Astrophysics Center, University of Wisconsin{\textemdash}Madison, Madison, WI 53706, USA \\
% ...
$^{41}$ Department of Physics, Marquette University, Milwaukee, WI 53201, USA \\
% ...
$^{43}$ Bartol Research Institute and Dept. of Physics and Astronomy, University of Delaware, Newark, DE 19716, USA \\
% ...
$^{63}$ Deutsches Elektronen-Synchrotron DESY, Platanenallee 6, D-15738 Zeuthen, Germany \\
$^{\rm a}$ also at Institute of Physics, Sachivalaya Marg, Sainik School Post, Bhubaneswar 751005, India \\
% ...

\subsection*{Acknowledgments}

\noindent
The authors gratefully acknowledge the support from the following agencies and institutions:
USA {\textendash} U.S. National Science Foundation-Office of Polar Programs,
U.S. National Science Foundation-Physics Division,
U.S. National Science Foundation-EPSCoR,
U.S. National Science Foundation-Office of Advanced Cyberinfrastructure,
Wisconsin Alumni Research Foundation,
Center for High Throughput Computing (CHTC) at the University of Wisconsin{\textendash}Madison,
Open Science Grid (OSG),
Partnership to Advance Throughput Computing (PATh),
Advanced Cyberinfrastructure Coordination Ecosystem: Services {\&} Support (ACCESS),
Frontera and Ranch computing project at the Texas Advanced Computing Center,
U.S. Department of Energy-National Energy Research Scientific Computing Center,
Particle astrophysics research computing center at the University of Maryland,
Institute for Cyber-Enabled Research at Michigan State University,
Astroparticle physics computational facility at Marquette University,
NVIDIA Corporation,
and Google Cloud Platform;
Belgium {\textendash} Funds for Scientific Research (FRS-FNRS and FWO),
FWO Odysseus and Big Science programmes,
and Belgian Federal Science Policy Office (Belspo);
Germany {\textendash} Bundesministerium f{\"u}r Forschung, Technologie und Raumfahrt (BMFTR),
Deutsche Forschungsgemeinschaft (DFG),
Helmholtz Alliance for Astroparticle Physics (HAP),
Initiative and Networking Fund of the Helmholtz Association,
Deutsches Elektronen Synchrotron (DESY),
and High Performance Computing cluster of the RWTH Aachen;
Sweden {\textendash} Swedish Research Council,
Swedish Polar Research Secretariat,
Swedish National Infrastructure for Computing (SNIC),
and Knut and Alice Wallenberg Foundation;
European Union {\textendash} EGI Advanced Computing for research;
Australia {\textendash} Australian Research Council;
Canada {\textendash} Natural Sciences and Engineering Research Council of Canada,
Calcul Qu{\'e}bec, Compute Ontario, Canada Foundation for Innovation, WestGrid, and Digital Research Alliance of Canada;
Denmark {\textendash} Villum Fonden, Carlsberg Foundation, and European Commission;
New Zealand {\textendash} Marsden Fund;
Japan {\textendash} Japan Society for Promotion of Science (JSPS)
and Institute for Global Prominent Research (IGPR) of Chiba University;
Korea {\textendash} National Research Foundation of Korea (NRF);
Switzerland {\textendash} Swiss National Science Foundation (SNSF).

\end{document}